\documentclass[5p]{elsarticle}

\usepackage{siunitx}
\usepackage{url}
\usepackage[]{graphicx}
\usepackage{amsmath}
\usepackage{amssymb}
\usepackage{bbm}
\usepackage{url}
\usepackage{booktabs}
\usepackage{multirow}
\usepackage{rotating}
\usepackage{mathtools}

\biboptions{longnamesfirst,sort&compress}

\usepackage{array}
\newcolumntype{L}[1]{>{\raggedright\let\newline\\\arraybackslash\hspace{0pt}}m{#1}}
\newcolumntype{C}[1]{>{\centering\let\newline\\\arraybackslash\hspace{0pt}}m{#1}}
\newcolumntype{R}[1]{>{\raggedleft\let\newline\\\arraybackslash\hspace{0pt}}m{#1}}

\newcommand{\Matrix}[1]{\ensuremath{\mathbf{#1}}}
\newcommand{\Vector}[1]{\ensuremath{\mathbf{#1}}}

\graphicspath{{figures/}}

\begin{document}

\begin{frontmatter}

\title{Compensating class imbalance for acoustic chimpanzee detection with convolutional recurrent neural networks}

\author[labp]{Franz Anders\corref{corautor}}
\ead{franz.anders@htwk-leipzig.de}
\author[uvic,mpi]{Ammie K. Kalan}
\ead{ammiek07@gmail.com}
\author[idiv,mpi]{Hjalmar S. K\"uhl}
\ead{hjalmar.kuehl@idiv.de}
\author[labp]{Mirco Fuchs}
\ead{mirco.fuchs@htwk-leipzig.de}

\address[labp]{Leipzig University of Applied Sciences, Laboratory for Biosignal Processing, Eilenburger Stra\ss{}e 13, 04317 Leipzig, Germany}

\address[uvic]{Department of Anthropology, University of Victoria, PO Box 1700 STN CSC, Victoria, BC, V8W 2Y2 Canada}

\address[idiv]{German Centre for Integrative Biodiversity Research (iDiv) Halle-Jena-Leipzig, Puschstra\ss{}e 4, 04103 Leipzig, Germany}

\address[mpi]{Max Planck Institute for Evolutionary Anthropology, Deutscher Platz 6, 04103 Leipzig, Germany}

\cortext[corautor]{Corresponding author}

\begin{abstract}

Automatic detection systems are important in passive acoustic monitoring (PAM) systems, as these record large amounts of audio data which are infeasible for humans to evaluate manually. In this paper we evaluated methods for compensating class imbalance for deep-learning based automatic detection of acoustic chimpanzee calls. The prevalence of chimpanzee calls in natural habitats is very rare, i.e. databases feature a heavy imbalance between background and target calls. Such imbalances can have negative effects on classifier performances. We employed a state-of-the-art detection approach based on convolutional recurrent neural networks (CRNNs). We extended the detection pipeline through various stages for compensating class imbalance. These included (1) spectrogram denoising, (2) alternative loss functions, and (3) resampling. Our key findings are: (1) spectrogram denoising operations significantly improved performance for both target classes, (2) standard binary cross entropy reached the highest performance, and (3) manipulating relative class imbalance through resampling either decreased or maintained performance depending on the target class. Finally, we reached detection performances of $\SI{33}{\percent}\ F1$ for drumming and $\SI{5}{\percent}\ F1$ for vocalization, which is a $>7$ fold increase compared to previously published results. We conclude that supporting the network to learn decoupling noise conditions from foreground classes is of primary importance for increasing performance.

\end{abstract}


\end{frontmatter}


\section{Introduction}

Automatic detection of chimpanzee calls is of primary importance for automatic monitoring of wild chimpanzee populations. There is a multitude of monitoring applications, from assessing chimpanzee home ranges for behavioral studies to early-warnings systems for areas with human-wildlife conflict.\cite{kalan2015towards, kalan2016passive}

Passive acoustic monitoring (PAM) is one of the most widely employed methods for monitoring wild animals. Here, autonomous recording units (ARUs) are distributed over an area for constant soundscape recording. The main advantages of PAM are: (1) minimal intrusion, as humans are only required for installation and maintenance of devices, (2) sampling over large spatial and temporal scales, and (3) detection of animals in habitats where visual recognition is limited, e.g. dense rain forests. However, PAM also produces large amounts of audio data which quickly become infeasible to curate manually by humans. Consequently, algorithms for automatic detection of target species are of primary importance for PAM settings.\cite{kalan2016passive}

Heinicke et al. \cite{heinicke2015assessing} investigated an automatic system for detecting calls of various primate species in PAM recordings of a tropical forest. Their algorithm employed a conventional acoustic detection approach based on hand-crafted features and gaussian mixture models. Algorithm performances varied with respect to target species and call type, from \SI{10}{\percent} $F1$ for Diana monkeys and King colobus monkeys to \SI{4}{\percent} for chimpanzee drumming and \SI{0.2}{\percent} for chimpanzee vocalizations. To the best of our knowledge, this is the only paper to date which focused on automatic detection of chimpanzee calls in a PAM setting. Dev's Master Thesis \cite{dev2020automatic} also investigated classification of chimpanzee calls, however for classification against the Urbansound8K  \cite{salamon2014dataset} dataset classes (e.g. car horn or gun shot) instead of detection in natural soundscape recordings.

In recent years, deep-learning based methods have become prevalent and have largely replaced approaches based on hand-crafted features in automatic audio recognition systems. In the annual DCASE-challenges, deep-learning based systems became popular between 2016 and 2017~\cite{mesaros2017detection, DCASE2017Task2Overview, DCASE2017Task1Overview, DCASE2019Overview}. Research teams working on automatic animal call detection particularly adapted convolutional neural networks (CNNs) with spectrogram inputs: Bergler et al. \cite{bergler2019orca} applied  Res-Net \cite{he2016deep} variants for detection of orca calls in long-term recordings; Bjorck et al. \cite{bjorck2019automatic} applied Dense-Net CNNs \cite{huang2017densely} for detecting African forest elephants with PAM; Oikarinen et al. \cite{oikarinen2019deep} applied siamese CNNs with stereo inputs to the detection of various marmoset monkey calls. Aodha et al. \cite{mac2018bat} investigated CNNs for bat detection. In an open challenge for bird audio detection in 2017\cite{stowell2018automatic}, the top placed system "bulbul" likewise applied CNNs to spectrogram inputs.

However, research is still lacking in some areas of automatic primate call detection as well as automatic animal call detection in general.

(1) Consideration of target class rarity in PAM settings. The majority of recordings in long-term PAM recordings will comprise background noise rather than target calls \cite{heinicke2015assessing, bjorck2019automatic,oikarinen2019deep}. Consequently, databases feature a heavy imbalance between background and target class. Numerous studies showed that class imbalances can have detrimental effects on automatic system performances, as classifiers are usually biased towards the majority class \cite{weiss2004mining, haixiang2017learning, johnson2019survey}. However, all previously mentioned studies worked with databases with strongly reduced amounts of background samples, biasing the class distribution towards the positive class. This bias was either already present in the respective database, or produced by the authors through discarding fixed percentages of noise samples. Additionally, nearly all papers measured system performances with metrics unsuited for unbalanced settings (accuracy or AUC ROC). These give overly optimistic results in recordings with heavy class imbalance as they are biased towards the majority class \cite{johnson2019survey}.

(2) Time-continuous detection. All previously mentioned deep-learning based systems approached detection tasks by classifying broader spectrogram patches of various seconds, e.g. 25-second patches \cite{bjorck2019automatic}. An evaluated temporal context receives a single class label. This approach decreases the temporal resolution of training annotations as well as test time predictions. In general-purpose detection tasks, convolutional recurrent neural networks (CRNNs) with time-distributed outputs have recently become more prevalent \cite{cakir2017convolutional}. They are capable of predicting each individual spectrogram time frame while incorporating broader spectrogram contexts. Such networks have not yet been applied to animal call detection tasks.

In this study, we investigated time-continuous detection of chimpanzee calls using CRNNs. The target calls were chimpanzee drumming and vocalizations in long-term PAM audio recordings of an African rain forest (Ta\"{i} National Park, C\^{o}te d'Ivoire). We addressed a particular challenge imposed by the severe imbalance between target classes and background samples in the database, caused by the rarity of chimpanzee calls. The test set, with a total duration of 179 hours, contained merely $\approx$ 10 minutes of chimpanzee calls. Consequently, we studied various methods for compensating this imbalance. These methods comprised (a) spectrogram denoising for compensating the absolute rarity of target classes, and (b) alternative loss functions and resampling (over \& undersampling). The novel contributions of this paper are:  

(1) the investigation of the feasibility of time-continuous detection with CRNNs for chimpanzee detection. To the best of our knowledge, this is the first application of this approach for animal species detection in long-term recordings.
(2) a dedicated investigation of various methods for compensating class imbalance, including resampling, alternative loss functions and spectrogram denoising.

\section{Materials and Methods}
\label{sec:methods}

\subsection{Dataset}
\label{sec:dataset}

The dataset was originally collected by AKK \cite{kalan2015towards, kalan2016passive} with the aim of developing an automated approach for detecting primate calls in PAM forest recordings \cite{heinicke2015assessing}. Heinicke et al. \cite{heinicke2015assessing} present the evaluation of this automated system.

The recording site was the western section of the Ta\"{i} National Park, C\^{o}te d'Ivoire. The area sampled the territories of two chimpanzee communities. The soundscape of the park featured a wide variety of biogenic sounds, e.g. birds, insects, anthropogenic sounds, rain, wind etc. The recording setup comprised 20 ARUs distributed evenly across an area of $\approx \SI{35}{\kilo\meter\squared}$. ARUs recorded in stereo with a sampling rate of \SI{16}{\kilo\hertz} and 16 bit depth. The recording period was from November 2011 to May 2012. ARUs recorded daily from 6 am to 6 pm on the full hour for 30 minutes. A total of \SI{12889}{\hour} of audio data was collected.

\begin{figure}[t]
    \centering
    \includegraphics[width=0.4\textwidth]{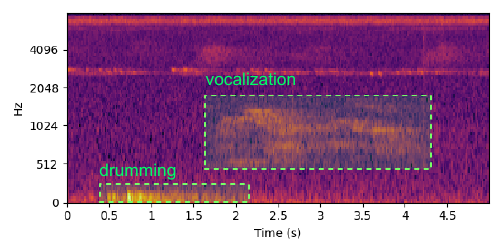}
    \caption{\textbf{Example spectrogram of target classes.}}
    \label{fig:example_spectrograms}
\end{figure}

\begin{figure}[t]
    \centering
    \includegraphics[width=0.4\textwidth]{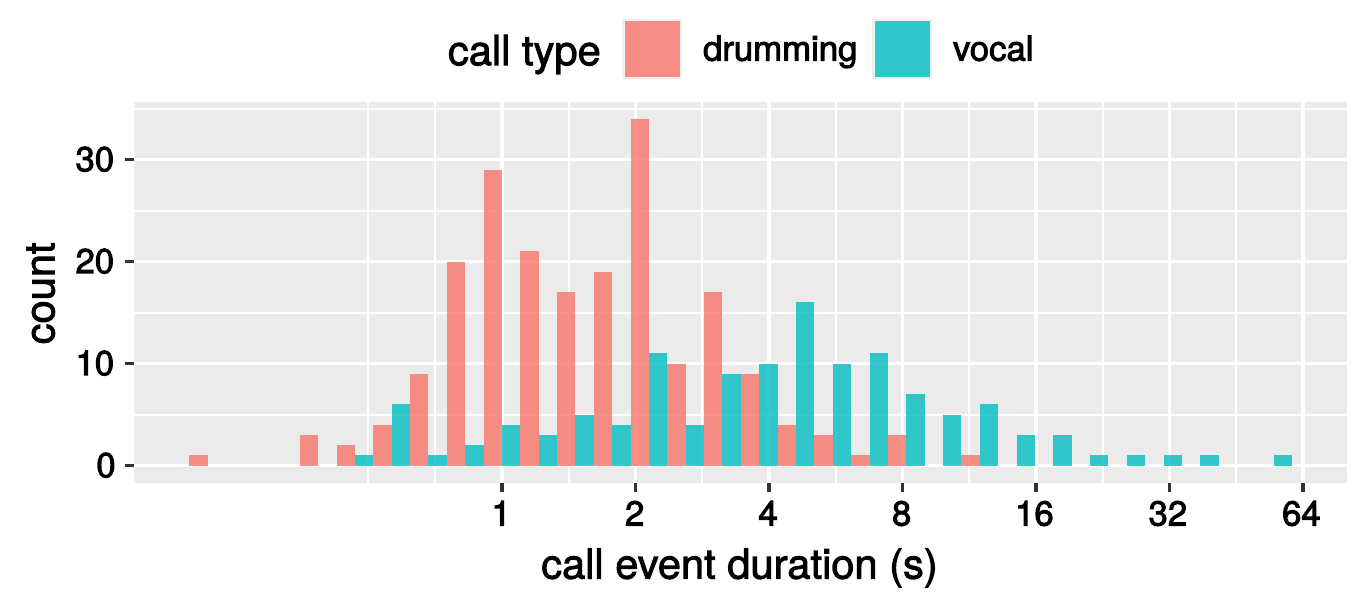}
    \caption{\textbf{Histogram of call type durations}}
    \label{fig:duration_histogram}
\end{figure}

The original automated approach \cite{heinicke2015assessing} targeted chimpanzees, \emph{Pan troglodytes} ssp. \emph{verus}, as well as three other primate species. The present study focuses exclusively on chimpanzees. Heinicke et al. defined two chimpanzee call types for detection: (1) \textbf{Drumming}, which is produced by chimpanzees when they repeatedly hit buttress roots of trees with their hands and feet. (2) \textbf{Vocalizations}, referring primarily to chimpanzee pant-hoots and screams for long-distance communication. Drumming is characterized by short energy bursts with low frequency. Vocalizations in the data set are characterized by harmonic patterns with an estimated frequency range of 200 - \SI{2000}{\hertz}. Figure \ref{fig:example_spectrograms} shows an example spectrogram excerpt with both call types. For the remainder of this paper, we refer to call types as \emph{classes} and call instances as \emph{events} in accordance with the vocabulary established in general audio detection research \cite{virtanen2018computational}.

To construct sets for training and validation of the automated system, the data pool was sampled, partitioned into sets of recordings, and annotated. Table \ref{tab:dataset} summarizes the datasets used in this study: The \emph{complete test set} corresponds to the original test set constructed by Heinicke et al. \cite{heinicke2015assessing}. They randomly sampled 358 recordings (of 30 minutes each) from the data pool, balanced across ARUs (one file per ARU per week) and time of day. This procedure ensured that the test set reflected diverse acoustic conditions for varying seasons, daytimes and sampling sites. We additionally constructed a \emph{reduced test set} which comprised all recordings from the complete test set with at least one chimpanzee event. We used the reduced test set for repeated evaluation runs, as the complete test set was computationally expensive due to its size. The \emph{training set} contained 44 additional recordings which were likewise randomly sampled (i.e. training and test set comprised of distinct recordings). Each individual recording in the training and test set was \SI{30}{\minute} long. The \emph{validation dataset} contained 25 additional recordings collected during a pilot study at the same location in 2010. Contrary to the other sets, recordings in the validation set had varying lengths with a mean duration of \SI{1.3}{\minute}. Two trained experts for primate vocalizations annotated call events in these recordings with precise start- and end times \cite{heinicke2015assessing}.

A central characteristic of the dataset is the rarity of the target classes. The complete test set with approximately one week of recording time merely contained a total of \SI{2.5}{\minute} of drumming and \SI{7}{\minute} of vocalization events. The relative amount for drumming and vocalizations was \SI{0.02}{\percent} and \SI{0.06}{\percent} respectively. This imbalance is representative of the real-world prevalence of chimpanzee calls obtained using PAM in natural settings. Even the reduced test set, which biased the class distribution in favor of the target classes, contained \SI{0.16}{\percent} and \SI{0.47}{\percent} of drumming and vocalization events respectively. The training set contained \SI{0.2}{\percent} and \SI{0.35}{\percent} of drumming and vocalization events respectively. 

We highlight that the imbalance between the number of positive class examples (i.e. target calls) and negative class examples (i.e. background samples) has two effects, according to the taxonomy of Weiss et al. \cite{weiss2004mining}: 
\begin{itemize}
\item \emph{Absolute rarity}, i.e. low amounts of training examples for the target classes. This causes classifiers to overfit individual examples rather than learning generalized patterns for the target classes, particularly in deep-learning systems \cite{virtanen2018computational, Goodfellow-et-al-2016}.

\item \emph{Relative imbalance} between background class and target class examples. This usually induces a prediction bias into the classifier to favor the majority class, while the minority class often is of greater interest to the user \cite{haixiang2017learning, johnson2019survey, weiss2004mining,leevy2018survey}. However, the magnitude of the negative effect depends on the complexity of the classification task at hand. The algorithm might be completely unaffected if classes are linearly separable \cite{japkowicz2000class, johnson2019survey}.
\end{itemize}

\begin{table}[t]
\renewcommand{\arraystretch}{1.1}
\centering
\footnotesize
\caption{\textbf{Dataset overview}}

\begin{tabular}{@{}L{0.6cm}L{1.7cm}L{1.2cm}L{1cm}L{1cm}L{1cm}@{}}
\toprule
               &                                       & test complete                       & test reduced                         & training                            & validation                          \\ \midrule
\multirow{3}{\linewidth}{ARU  record.} 	& \# recordings                         & 358                                 & 50                                   & 44                                  & 25                                  \\
               & total duration                & $\SI{179}{\hour}$                                 & \SI{25}{\hour}                                   & \SI{22}{\hour}                                  & \SI{0.7}{\hour}                                 \\  
               & \% recordings with at least 1 chimp. call & $\SI{14}{\percent}$ &  $\SI{100}{\percent}$ &  $\SI{50}{\percent}$ & $\SI{76}{\percent}$ \\ \midrule
\multirow{3}{\linewidth}{drum.}     & \# events                             & 100                                 & 100                         & 78                                  & 29                                  \\
               & total duration            & \SI{149}{\second}                                 & \SI{149}{\second}                        & \SI{159}{\second}                                 & \SI{96}{\second}\\
               & mean duration              & \SI{1.29}{\second}                                  &   \SI{96}{\second}                        & \SI{1.76}{\second}                                  & \SI{2}{\second}                                     \\  \midrule
\multirow{3}{\linewidth}{vocal.}    	& \# events                             & 50                                  & 50                         & 53                                  & 23                                  \\
               & total duration              &  \SI{431}{\second}                                 & \SI{431}{\second}                         &  \SI{270}{\second}                                 &  \SI{88}{\second}                                  \\
               & mean duration             &  \SI{5.72}{\second}                                & \SI{5.72}{\second}                         &  \SI{4.35}{\second}                                &  \SI{3}{\second}                                   \\ \bottomrule
\end{tabular}
\label{tab:dataset}
\end{table}

\subsection{Detection pipeline}
\label{sec:approach_overview}

\begin{figure}[t]
    \centering
    \includegraphics[width=0.5\textwidth]{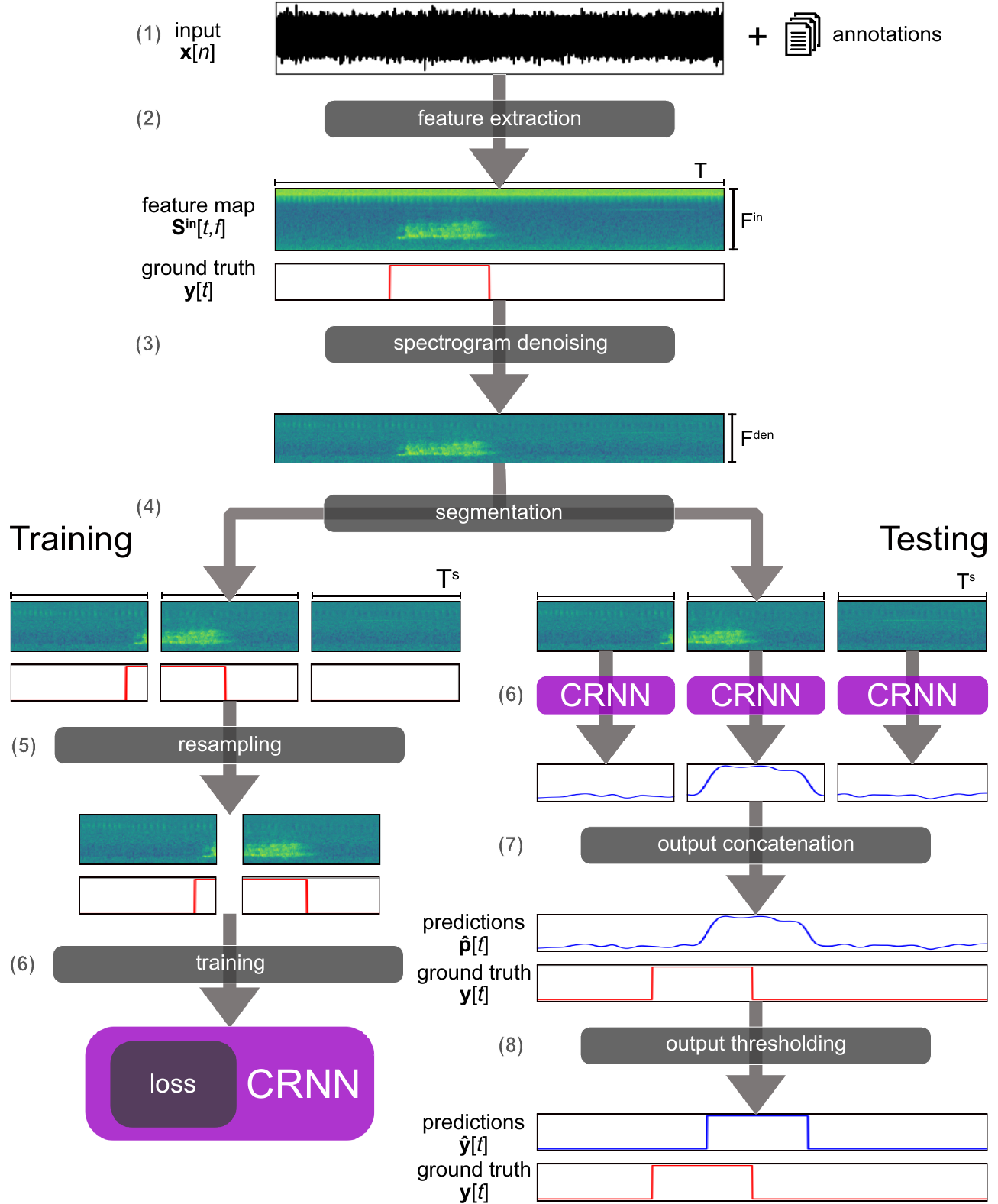}
    \caption{\textbf{Overview of the detection pipeline at training and test time}. In this example there is only one input signal $\Vector{x}$, so that the index $i$ is omitted. }
    \label{fig:pipeline_overview}
\end{figure}

Figure \ref{fig:pipeline_overview} gives an overview of the detection pipeline at training and test time. The pipeline consist of a series of \emph{pipeline stages} which progressively process an input audio signal (i.e. ARU recording). At training time, the pipeline outputs a trained CRNN. At test time, the pipeline outputs indications on target class presence for spectrogram time frames. 

The general approach and stage arrangement of \emph{feature extraction}, \emph{segmentation}, \emph{CRNN}, \emph{output concatenation} and \emph{output thresholding} originates from Cakir et al.\cite{cakir2017convolutional}. We additionally added the stages \emph{spectrogram denoising} and \emph{resampling}. These stages, together with the choice of the loss function, are the three components aimed at compensating class imbalance investigated in this study. Although Cakir's pipeline is aimed at polyphonic detection tasks, we only considered single class detection in this study, i.e. a separate CRNN must be trained for drumming and vocalization. This restriction was imposed to study the effects for both classes individually.

The pipeline stages function as follows:

(1) \emph{input}: 
Input are ARU recordings $\Vector{x}_i \in \mathbb{R}^{l_i}$ of length $l_i\in \mathbb{N}$ as time-domain audio signals, where $i$ is the signal index. Signals are converted to mono and normalized to a peak amplitude of 1 to equalize loudness across signals. Ground truth annotations are tables which list occurrences of target calls with their start- and end times within signals.

(2) \emph{feature extraction}: 
Signals $\Vector{x}_i$ are converted into spectrograms $\Matrix{S}^{\text{in}}_i \in \mathbb{R} ^{T_i \times F^{\text{in}}}$, where $T_i \in \mathbb{N}$ is the respective amount of spectrogram time frames and $F^{\text{in}} \in \mathbb{N}$ is the amount of frequency bins. We employed mel-scaled spectrograms as it currently is the most prevalent choice in deep-learning based sound recognition systems ~\cite{hershey2017cnn, DCASE2017Task2Overview, DCASE2017Task1Overview, DCASE2019Overview}. We adapted the variant of the DCASE baseline systems~\cite{mesaros2018multi}, which is the logarithm of a slaney-stile mel-fitlerbank of the linear magnitude spectrogram. We adapted the frame length of \SI{40}{\milli\second} and hop-length of \SI{20}{\milli\second} from Cakir et al. We used $F^{\text{in}} = 80$ mel-bands, i.e. we doubled Cakir's frequency resolution to reduce the loss of potentially important frequency information, particularly in view of the low signal-to-noise-ratio in this task. The feature extraction process was implemented through the python library \texttt{librosa v.0.8.}~\cite{mcfee2015librosa}.
Ground truth annotations are likewise converted to binary target vectors $\Vector{y}_i\in \{0,1\}^{T_i}$, where $\Vector{y}_i[t]=1 $ encodes presence and $0$ encodes absence of the target class in time frame $t$ (alias "positive" and "negative" example).

(3) \emph{spectrogram denoising}:
This stage applies a series of denoising functions to spectrograms $f^{\text{den}} (\Matrix{S}^{\text{in}}_i) = \Matrix{S}^{\text{den}}_i \in \mathbb{R}^{T_i \times F^{\text{den}}}$. While the time axis size $T_i$ remains unchanged, the frequency axis might be reduced to $F^{\text{den}}$. This stage also z-standardizes all spectrograms through global statistics calculated on the training set.

(4) \emph{segmentation}:
While all spectrograms $\Matrix{S}^{\text{den}}_i $ contain the same amount of frequency bands, the number of time frames $T_i$ might vary depending on the input duration for signal $i$. However the CRNN requires inputs with equally sized dimensions to be trained via mini batch training. Therefore spectrograms are segmented across the temporal axis into non-overlapping segments of fixed temporal length $T^{\text{seg}}$. Consequently, $\Matrix{S}^{\text{den}}_i$ is represented as a list of segments $\Matrix{S}_{i,j} \in \mathbb{R} ^{T^{\text{seg}} \times F^{\text{den}}}$, where $j$ is the segment index. For training, target vectors are likewise segmented into $\Matrix{y}_{i,j} \in \mathbb{R} ^{T^{\text{seg}}}$. In this study, we used a fixed segment length of $T_s \hat{=} \SI{10}{\second}$ as a balance between computational efficiency and providing sufficient temporal context in spectrogram segments for recognition of target classes. For signals with $T_i\ \text{mod}\ T^{seg} \neq 0$ (some signals in the validation set), the last segment was partially overlapped with the penultimate to cover $T_i$ completely if the overlap was $< \SI{75}{\percent}$, or discarded otherwise.

(5) \emph{resampling}: This stage alters the distribution between positive and negative examples through over and undersampling of segments. This stage is only active for training to not affect class distribution when validating the system. Section \ref{sec:resampling} provides details on the resampling procedure. 

(6) \emph{CRNN prediction / training}: 
The CRNN processes spectrogram segments individually to produce corresponding class probability vectors $f^{\text{crnn}}(\Matrix{S}_{i,j}) = \hat{\Vector{p}}_{i,j} \in [0,1] ^{T_s}$, i.e. $\hat{\Vector{p}}_{i,j}[t]$ indicates the probability of the target class being present in frame $t$ of segment $j$ for signal $i$. At training time, predictions are used for iterative optimization of network weights. Section \ref{sec:crnn_architecture} provides details on the CRNN configuration. We experimented with various loss functions as described in section \ref{sec:loss}.

(7) \emph{output concatenation}: The stage concatenates the prediction segments to produce one target vector per input spectrogram $[\Vector{\hat{p}}_{i,0}, \Vector{\hat{p}}_{i,1}, ...] = \Vector{\hat{p}}_{i} \in \mathbb{R} ^{T_i}$.

(8) \emph{output binarization}: This stage binarizes predicted probabilities through thresholding: $\Vector{\hat{y}}_{i} = 1$, if $\Vector{p}_{i} > C$, else 0. We used the unbiased threshold $C = 0.5$ as in \cite{cakir2017convolutional}.

\subsection{Spectrogram denoising}
\label{sec:denoising}

Spectrogram denoising is a common preprocessing step in automatic animal call detection. The aim is to increase the signal-to-noise-ratio, as recordings usually carry high amounts of noise due to the nature of open field settings. In this context, the signal is the target call of interest. Noise refers to \emph{background sounds}, i.e. geophony (environmental sounds such as wind and rain), anthrophony (noise generated by humans, such as traffic) and biophony (sounds of animals not of interest). In the context of this work, we employed spectrogram denoising as a method for combating absolute rarity of target classes. Target event examples are present for only few background noise conditions, possibly causing the classifier to infer false coupling between noise conditions and event probability. Prior elimination of variability between noise conditions can mitigate this effect.  \cite{xie2020bioacoustic, mac2018bat, himawan2018deep}

Among the multitude of available methods, we chose to evaluate: (1) frequency removal and (2) spectral subtraction, as they are among the most prevalent and straight-forward methods in automatic animal call detection \cite{xie2020bioacoustic, mac2018bat, himawan2018deep}. Both methods exploit the observation that background noise is fairly consistent over large periods of time, while target classes are comparatively short and seldom.

\subsubsection{Frequency removal}
\label{sec:frequency_removal}

Animal target calls usually occupy narrow frequency ranges. Therefore, many systems apply preprocessing filters to remove unneeded frequency ranges \cite{haixiang2017learning}. We calculated frequency ranges for target classes through the following proposed method:

The goal is to calculate a \emph{class mask} $\Vector{r}\in \mathbb{R}^{F}$, whose values $\Vector{r}[f]$ indicate the strength of association between mel frequency bins $f$ and the target class. let $T^{\text{start}}_{e} \in \mathbb{N}$ and $T^{\text{end}}_{e} \in \mathbb{N}$ be the start and end time of event $e \in \{0, ..., E-1\}$ indicated as spectrogram time frame indices. $\Matrix{S}_e \in \mathbb{R}^{T \times F}$ and $\Vector{y}_e$ are the spectrogram and ground truth vectors which contain event $e$ at some points. $t_e \in [T^{\text{start}}_{e} - T^c , ... ,  T^{\text{end}}_{e} + T^c]$ is the list of time frames for event $e$, padded by a fixed context size $T^c$. Then, $\Matrix{S}^{\text{context}}_e$ and $\Matrix{y}^{\text{context}}_e$ are the spectrogram and ground truth vector patches corresponding to $t_e$, i.e. $\Matrix{S}^{\text{context}}_e = \{\Matrix{S}_e[t,f]\ |\ t \in t_e \}$. For each event, we obtain an \emph{event mask} $\Vector{r}_e\in \mathbb{R}^{F}$ by calculating the Pearson correlation between the mel frequency bin $f$ and the target vector as:

\begin{equation}
 \Vector{r}_{e}[f] = \frac{\text{cov}(\Matrix{S}^{\text{context}}_e[f], \Matrix{y}^{\text{context}}_e[f] )}
{\sigma ( \Matrix{S}^{\text{context}}_e[f]), \sigma (\Matrix{y}^{\text{context}}_e[f] ) }
\end{equation}

The total class mask $\Vector{r}$ is the average of all event masks:
\begin{equation}
 \Vector{r}[f] = \frac{1}{E} \sum_{e=0}^{E-1}\Vector{r}_{e}[f]
\end{equation}

The intuition behind this calculation method is as follows: Any call event causes an increase in energy of its associated frequency bands relative to the background noise. If an event is surrounded by time invariant noise, this gain is measurable as a positive correlation between a frequency band's energy and the target vector. The class mask value range is $ -1 \leq \Vector{r}[f] \leq +1$, where $+1 / -1$ indicates perfect association of a frequency band to the target vector and $0$ indicates no association. Positive correlations $>0$ are attributed to the actual target class acoustic content, i.e. energies active during target events. Negative correlations $<0$ indicate systematic absence of band energies during events, which might be caused by other sounds commonly preceding, succeeding or pausing during target events. Our approach is applicable for calculating frequency ranges of arbitrary sound classes, if the mentioned preconditions are met.

\begin{figure}[t]
    \centering
    \includegraphics[width=0.35\textwidth]{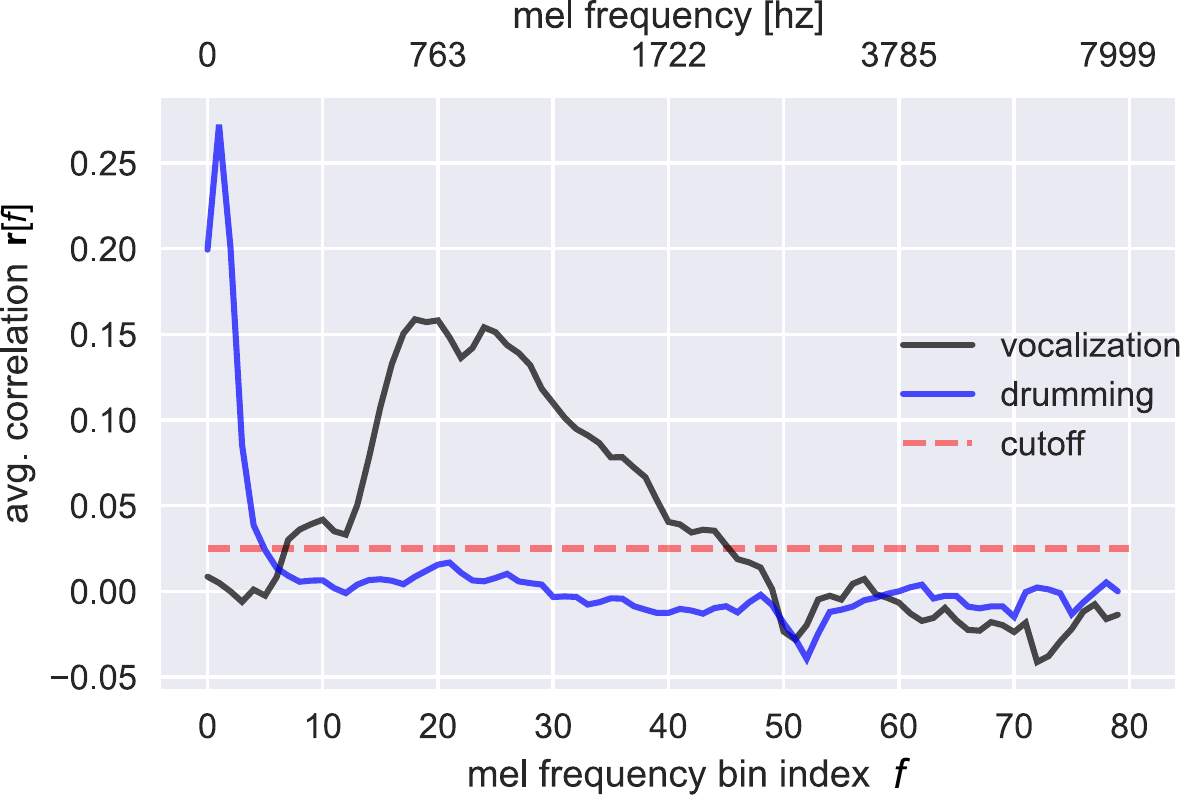}
    \caption{\textbf{Class masks.} Graphs indicate the strength of association between mel frequency bands and target classes. $\Vector{r}[f] = 0$ indicates no association and $\Vector{r}[f] = +1 / -1$ indicates perfect association. The red line indicates the threshold under which we removed frequency bands when applying frequency removal.}
    \label{fig:class_masks}
\end{figure}

Figure \ref{fig:class_masks} shows the class masks for both target classes. These masks were calculated exclusively on events from the training and validation set. The context length was $T^c \hat{=} \SI{1}{\minute}$ which roughly corresponds to the length of the longest call event in our database. When applying frequency removal, we set a correlation threshold $C^r$ to remove frequency bands with $\Vector{r}[f] < C^r$. We chose $C^r = 0.025$ as a rather low threshold to ensure no significant target class information was lost. We exclusively considered positive correlations as only those were indicative of the actual target class' signal content. Negative correlations, particularly $> \SI{2500}{\hertz}$, were attributed to variations in the background noise conditions such as periodic insect calls which we aimed to eliminate in this preprocessing step.

This procedure retained the lowest 5 frequency bins for drumming (range $0 - \SI{152}{\hertz}$), and 39 frequency bins for vocalization (range $267 - \SI{2097}{\hertz}$). Those frequency ranges correspond to estimations of previous studies \cite{heinicke2015assessing}. Figure \ref{fig:preproc_example} visualizes the effect of frequency removal.

\subsubsection{Spectral subtraction}
\label{sec:spectral_subtraction}

Spectral subtraction removes background noise by subtracting a noise profile from signals. If the noise profile is assumed invariant across time/recordings, the profile is commonly estimated by averaging each frequency bin across all time steps inside a background noise region \cite{xie2020bioacoustic}. However, we observed that noise conditions vary strongly between recordings due to time of day, season and ARU location, but are fairly consistent within recordings. Consequently, we estimated local noise profiles for recordings as follows:

A spectrogram $\Vector{S}_i$ is segmented into non-overlapping segments $\Vector{S}_{i,j} \in \mathbb{R}^{T^{\text{sub}} \times F}$ of length $T^{\text{sub}}$ similar to the segmentation step in the processing pipeline. From each segment we subtract each frequency bin's average value:

\begin{equation}
S^{\text{denoised}}_{i,j}[t,f] = \Matrix{S}_{i,j}[t,f] - \frac{1}{T^{\text{sub}}} \sum^{T^{\text{sub}}}_{t=0} \Matrix{S}_{i,j}[t,f] 
\end{equation}

\begin{figure}[t]
    \centering
    \includegraphics[width=0.35\textwidth]{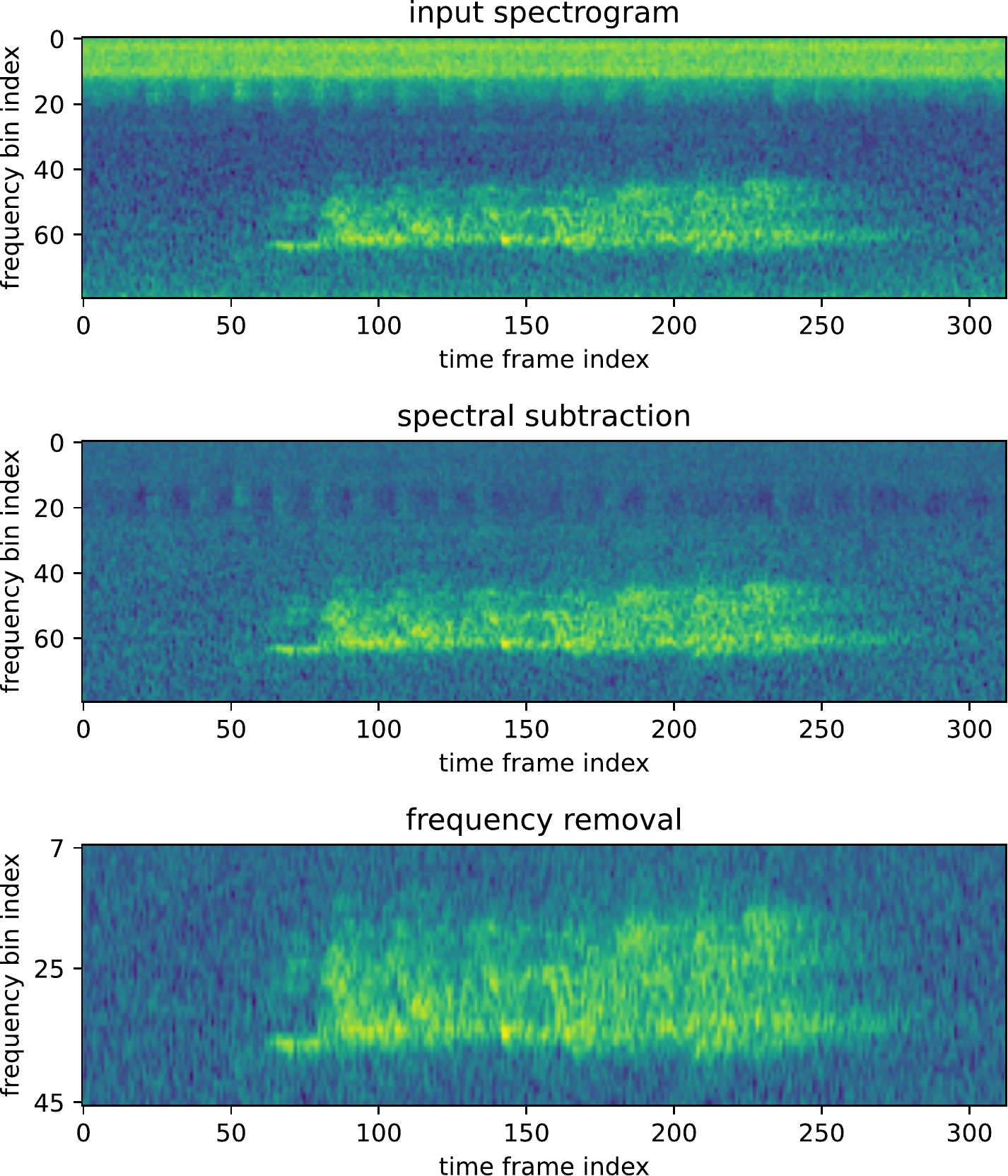}
    \caption{\textbf{Visualization of preprocessing operations for an example vocalization event.} Top: Input spectrogram. Middle: Effect of spectral subtraction (see sec. \ref{sec:spectral_subtraction}). Bottom: Effect of frequency removal (see sec. \ref{sec:frequency_removal}) }
    \label{fig:preproc_example}
\end{figure}

If $T^{\text{sub}}$ is sufficiently large compared to the expected target event duration, the average frequency band energy is dominated by the noise profile rather than the target class profile. Denoised spectrogram segments are then concatenated to reconstruct the input spectrogram progression. This method is similar to the one applied in study \cite{mac2018bat}. Here we set $T^{\text{sub}} \hat{=} \SI{3}{\minute}$, so that even the maximum expected call duration contains twice as much noise as the target signal. Figure \ref{fig:preproc_example} visualizes the effect of spectral subtraction.

\subsection{Convolutional Recurrent Neural Network}
\label{sec:crnn_architecture}

\begin{figure}[t]
    \centering
    \includegraphics[width=0.4\textwidth]{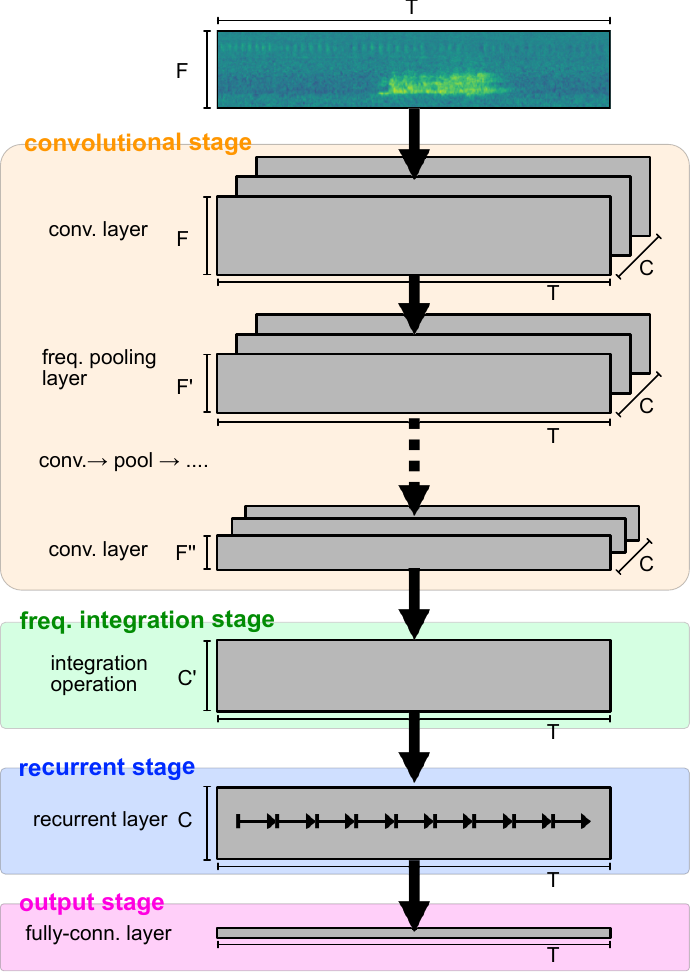}
    \caption{\textbf{Convolutional recurrent neural network architecture scheme.}}
    \label{fig:network_architecture}
\end{figure}

Figure \ref{fig:network_architecture} visualizes the network architecture scheme. The scheme is largely adapted from Cakir et al. \cite{cakir2017convolutional}. It contains a sequence of network stages which process a spectrogram $\Matrix{S}_{i,j} \in \mathbb{R} ^{T^{\text{seg}} \times F^{\text{den}}}$ to produce an output prediction vector $\hat{\Vector{p}}_{i,j} \in [0,1] ^{T^{\text{seg}}}$. For simplicity, we denote these elements as $\Vector{S} \in \mathbb{R} ^{T \times F}$ and $\hat{\Vector{p}} \in \mathbb{R}^{T}$ in this section. A central network property is that the temporal dimension $T$ remains intact throughout the network, i.e. the temporal alignment between spectrogram frames and output time steps is maintained.

\subsubsection{Architecture Scheme}

The \emph{input stage} appends an empty channel dimension to the spectrogram in preparation for the convolutional stage $\mathbb{R} ^{T \times F} \mapsto \in \mathbb{R} ^{T \times F \times (C=1)}$, 

The \emph{convolutional stage} consists of alternating convolutional and pooling layers similar to conventional CNN architectures \cite{krizhevsky2012imagenet}. Convolutional layers extract local features via learned filter kernels. They retain the size of the frequency and time dimension of their respective inputs through padding, but alter the channel dimension according to the number of filter kernerls $k$, i.e. $\text{conv}^k: \mathbb{R} ^{T \times F \times C} \mapsto \mathbb{R} ^{T \times F \times C' = k}$. Pooling layers reduce the frequency axis size through one-dimensional pooling according to the pooling stride $p$, i.e. $\text{pool}^p: \mathbb{R} ^{T \times F \times C} \mapsto \mathbb{R} ^{T \times (F' = F / p) \times C}$. The stage outputs a volume $\in \mathbb{R} ^{T \times F^{\text{conv}} \times C^{\text{conv}}}$, where $F^{\text{conv}}$ is the frequency size remaining after several pooling operations and $C^{\text{conv}}$ is the number of filter kernels in the last convolutional layer.

The \emph{frequency integration stage} removes the frequency dimension by incorporating frequency into channel information : $\mathbb{R} ^{T \times F^{\text{conv}} \times C^{\text{conv}}} \mapsto \mathbb{R} ^{T \times C^{\text{freqint}}}$. This is done via a time-distributed function, i.e. a function which is applied equally at each time step $t$.

The \emph{recurrent stage} contains a recurrent layer which processes the output of the frequency integration stage sequentially at every time step. The output of the recurrent layer at time step $t$ depends on the input of the preceding layer at the same time step $t$ as well as the hidden states of the recurrent layer at steps $0, ..., t-1$. The volume dimensionality transformation is $\mathbb{R} ^{T \times C^{\text{freqint}}} \mapsto \mathbb{R} ^{T \times C^{\text{rec}}}$. 

Finally, the \emph{output layer} consists of a time-distributed fully-connected layer $\mathbb{R} ^{T \times C^{\text{rec}}} \mapsto \mathbb{R} ^{T}$, i.e. a single fully-connected neuron which is applied with the same weights at each time step. We initialized the bias parameter of this neuron to the respective training class distribution $\log_e(pos/neg)$, where $pos$ and $neg$ are the amount of positive and negative time steps respectively. This way, networks start training with appropriate output distributions which can prevent instability in the initial training steps, as recommended by Lin et al. \cite{lin2017focal}

The central difference between Cakir's scheme and ours is the definition of the frequency integration stage as a generic stage, while Cakir defined a single fixed integration operation for flattening the frequency axis. The reason for this change is that we identified the frequency integration operation as one of the most important architecture hyper parameters in our previous studies \cite{anders2020automatic, anders2020comparison}. We implemented networks and training with \texttt{tensorflow v.2.3.1}.

\subsubsection{Network Configuration}

The following network parameters were fixed. For the convolutional stage, all convolutional layers used kernel sizes $(5,5)$ and strides of $(1,1)$. Convolutional layers were always followed by a batch-normalization layer and ReLU activation. Pooling layers always used max-pooling. The recurrent stage contained a single recurrent layer with gated recurrent units (GRU). Networks were trained with Adam optimizer with standard parameters \cite{kingma2014adam} and early-stopping based on the validation set loss. All of these parameters were chosen in accordance to the recommendations of Cakir et al. \cite{cakir2017convolutional}.

\begin{table}[t]
\renewcommand{\arraystretch}{1.2}
\footnotesize
\centering
\caption{\textbf{Search space for network configuration and spectrogram denoising.} Table header shows input class, the top half shows denoising setting, and the bottom half shows the search space for network configuration parameters. The symbol $\leftarrow$ indicates that a search space corresponds to the respective left cell. Abbreviations: GAP = 1d global average pooling, GMP = 1d global max pooling}

\begin{tabular}{@{}L{2.3cm}| L{1.5cm} L{1.2cm} | L{1.2cm}L{1.2cm}@{}}
\toprule
input class             & \multicolumn{2}{c}{vocal}                & \multicolumn{2}{c}{drumming}     \\ \midrule
freq. removal           & no                     & yes             & no              & yes            \\
input size $T \times F$ & $500 \times 80$        & $500 \times 39$ & $500 \times 80$ & $500 \times 5$ \\
spectrl. subtr.         & $\{no, yes\}$          & $\{no, yes\}$   & $\{no, yes\}$   & $\{no, yes\}$  \\  \midrule
\# units / layer      & $\{ 32, 64, 96 \}$     & $\leftarrow$    & $\leftarrow$    & $\leftarrow$   \\
\# conv. layers         & $\{ 2, 3 ,4 \}$        & $\leftarrow$    & $\leftarrow$    & $\{ 1, 2 \}$   \\
pool size \& stride        & $\{ 2, 3 ,4, 5 \}$     & $\leftarrow$    & $\leftarrow$    & $\{ 2, 3 \}$   \\
freq. integr. op.       & \{flatten, GAP, GMP \} & $\leftarrow$    & $\leftarrow$    & $\leftarrow$   \\ \bottomrule
\end{tabular}
\label{tab:network_parameters}
\end{table}

The following network parameters were subjected to a parameter search as part of the experimental setup in this study (see section \ref{sec:experimentation}). The \emph{channel size} $\in \mathbb{N}$ is the number of filters in each convolutional or recurrent layer, i.e. the number of filters/units is held constant across the entire network. The \emph{conv. stage depth} $\in \mathbb{N}$ determines the number of convolutional layers. We chose to search these parameters as they primarily determine the amount of network weights and consequently the capacity to fit the data \cite{Goodfellow-et-al-2016}. The \emph{pooling size} $\in \mathbb{N}$ determines the pooling size and stride for all pooling layers. The \emph{frequency integration operation} $\in \mathbb{R} ^{F \times C} \mapsto \mathbb{R} ^{C'}$ is the time-distributed operation for integration of the frequency dimension. We searched for these parameters as we found them to be of primary importance for generalization capability in prior studies \cite{anders2020automatic, anders2020comparison}. Finally, \emph{recurrent bi-directional} $\in \mathbb{B}$ determined whether the recurrent layer was used bi-directionally. If yes, half of the units run forward and half backwards. We searched this parameter as we hypothesized it to be of primary importance for the network's capability to precisely locate frames of event activity.

Table \ref{tab:network_parameters} summarizes the search space. As explained further in section \ref{sec:experimentation}, we investigated network configuration in combination with the spectrogram denoising stage setting. All search spaces were equal regardless of input class and denoising setting, except for drumming when frequency removal was applied, as this removed \SI{93}{\percent} of the input frequency axis size.

\subsection{Loss}
\label{sec:loss}

The choice of loss function is among the primary algorithm-level methods for mitigating class imbalance in neural networks \cite{johnson2019survey}. The standard loss for classification problems with output neurons with sigmoid activations is binary cross entropy \cite{Goodfellow-et-al-2016, cakir2017convolutional}: 

\begin{equation}
BCE(\hat{p}, y) = 
\begin{cases*}
- w_1 \log(\hat{p}) & \text{, if}\ y = 1\\
- w_0 \log(1-\hat{p})& \text{, else}
\end{cases*}
\end{equation}

where $w_1, w_0$ are optional weights for the background and the target class. In standard binary cross entropy both classes are weighted equally, i.e. $w_1, w_0 = 1$. Binary cross entropy was the default loss used for network optimization. As part of the experimental setup (see section \ref{sec:experimentation}), we additionally evaluated a selection of the most prevalent BCE variants \cite{johnson2019survey} aimed at mitigating class imbalance. These were:
\begin{itemize}
\item \textbf{weighted binary cross entropy} weighs classes according to the relative count of examples: $w_k= |C_k|/(|K| \cdot |C|)$, where $k \in K = \{0, 1\}$ is the class index, $C_k$ is the set of examples for class $k$, and $C$ is the total set of examples. This calculation method originates from King et al. \cite{king2001logistic}, as implemented with \texttt{scikit-learn v.0.23}.
\item \textbf{focal loss} down-weighs the loss of well-classified examples irregardless of the class trough: $w_0 = (\hat{y})^\gamma $ and $w_1 = (1-\hat{y})^\gamma$. The hyper parameter $\gamma$ determines the amount of down-weighting and is set to a default value $\gamma = 2$ in the paper \cite{lin2017focal}. We used the implementation of \texttt{tensorflow-addons 0.11.2}.
\item \textbf{weighted focal loss} is the combination of focal loss with class weights as previously described. 
\end{itemize}

\begin{table}[t]
\renewcommand{\arraystretch}{1.1}
\footnotesize
\center

\caption{\textbf{Search space for loss and resampling}}

\begin{tabular}{@{}L{1.3cm} L{2.2cm} |L{1cm}L{3cm}@{}}
\toprule
stage      & component                      & default & search space                       \\ \midrule
training   & loss                           & BCE           & \{weighted BCE, FL,  weighted FL\} \\
\multirow{2}{*}{resampling}  & oversampling dupl. amount      & 2             & $\{0, 2, 4,8, 16\}$                  \\
           & undersampling disc. percentage & 75\%          & \{0\%, 50\% 75\%, 90\%, 95\%\}    \\ \bottomrule
\end{tabular}
\label{tab:stages_searchspace}
\end{table}

\subsection{Resampling}
\label{sec:resampling}

Resampling is the most prevalent data-level technique for mitigating relative class imbalance. Resampling seeks to rebalance the distribution between class examples by undersampling, i.e. discarding examples from the majority class, and oversampling, i.e. duplicating examples from the majority class. Both techniques have been successfully applied for deep-learning based systems in imbalanced settings \cite{johnson2019survey}. Among the set of available resampling techniques, we chose to experiment with the most prevalent and straight-forward implementations: random over and undersampling.

We implemented resampling as follows. After the segmentation step, each ARU recording was represented as a list of spectrograms $\Matrix{S}_{i,j}$ and target vectors $\Matrix{y}_{i,j}$ where $i$ is the signal and $j$ is the segment index. We separated each list into disjoint subsets, those containing "negative" segments, i.e. segments with exclusively background time steps $Neg_i = \{j\  |\ \max(\Vector{y}_{i,j}) = 0)  \}$, and those containing "positive" segments, i.e. segments with at least one time step with a target call $Pos_i = \{ j\  |\ \max(\Vector{y}_{i,j}) = 1)  \}$.

The strength of undersampling is determined through a parameter $U \in [0,...,1]$ which indicates the percentage of discarded negative examples. The strength of oversampling is determined through a parameter $O \in \mathbb{N}$ which indicates the number of duplications of all positive examples, e.g. $O = 1$ means that each positive example is duplicated once. Undersampling and oversampling was performed for each input signal separately. Using this approach, we ensured that the set of background examples displayed a certain degree of diversity even for higher discarding percentages, as noise had greater variance between than within signals.

Table \ref{tab:stages_searchspace} shows the default settings and search space for the resampling stage. As shown, the default setting applies some amount of over- and undersampling, while not completely balancing distributions. The reasons were (a) decreasing training time for the initial experiments by reducing the data set size, (b) ensuring that the extreme imbalance in the training set does not prevent training convergence \cite{johnson2019survey}, and (c) to imitate the default setting commonly used in animal call detection systems, which usually start with already undersampled databases \cite{bjorck2019automatic, oikarinen2019deep, bergler2019orca}.

\subsection{Evaluation}

We chose the following metrics for evaluation of system performance: (1) average precision ($AP$), also known as the "area under the precision-recall-curve". This metric is based on the un-binarized prediction probabilities, i.e. comparing $\hat{\Vector{p}}$ and $\Vector{y}$ after the concatenation step described in section \ref{sec:approach_overview} (predictions / ground truth vectors for all signals were concatenated ). (2) $F1$, calculated on the binarized prediction probabilities $\hat{\Vector{y}}$ and $\Vector{y}$. Both metrics are recommended for the evaluation of imbalanced class distributions \cite{bjorck2019automatic, davis2006relationship}. All metrics were implemented with \texttt{scikit-learn v.0.23} 

We chose the \emph{segment based} approach as the evaluation method, i.e. metrics were based on comparing fixed-length time intervals as evaluation instances \cite{mesaros2016metrics}. We chose the following temporal resolutions:

\begin{itemize} 
\item Frame-wise resolution: Evaluation segments corresponded directly to spectrogram frame indications $\hat{\Vector{p}}$ or $\hat{\Vector{y}}$ and $\Vector{y}$. Consequently, the evaluation segment length was \SI{20}{\milli\second}. This resolution measures the system capacity for precise localization of events. It implicitly weighs target events according to the amount of time frames, i.e. longer events influence metric scores more than short events. This resolution was noted through the subscript $AP_{\text{frm}}$ and $F1_{\text{frm}}$.
\item 5 second resolution: Ground truth and prediction vectors were down-sampled to \SI{5}{\second} intervals as evaluation instances. The down-sampling was performed through max-pooling in non-overlapping segments of \SI{5}{\second} length. This resolution measures the system capacity for coarse localization of events. It also compensates the effect of event importance being weighted by length, as all events with length $< \SI{5}{\second}$ contribute the same amount of evaluation segments. However, it also over-penalizes short false-positive-peaks. This resolution was noted through the $AP_{\text{5}}$ and $F1_{\text{5}}$.
\item Averaged resolution: The average of both resolutions, e.g. $ F1_{\text{avg}} = (F1_{\text{frm}} + F1_{5}) / 2$
\end{itemize}

We used $AP_{\text{avg}}$ as the primary evaluation metric for the following reasons. (1) It is independent of a binarization threshold, which imposes another hyperparameter which might require tuning. (2) End users for this application prefer unbinarized predictions as being more informative and (3) It summarizes the system capacity for precise and coarse event localization. We indicated all metrics in the range $[0,...,1]$, where $0$ is the worst and $1$ is a perfect score.

\subsection{Experimental Setup}
\label{sec:experimentation}

The central goal of the experimenation was to evaluate the influence of pipeline stages for mitigating class rarity. Tables \ref{tab:network_parameters} and  \ref{tab:stages_searchspace} summarize the stage's search spaces and default settings. However, the search space was too large to exhaustively evaluate via a full-factorial design, i.e. evaluating each possible combination of parameters ("grid search"). Therefore, we split the experiment into two rounds to investigate local combinations of stages full-factorially:

\textbf{Round 1: Optimization of network architecture + spectrogram denoising}. For each setting of the denoising stage (no preprocessing / freq. removal / spec. sub. /  freq. removal + spec. sub.) we performed a full architecture grid search according to the search space shown in Table \ref{tab:network_parameters}. The goal was to identify the optimal combination of denoising setting and network architecture configuration. 
As this round performed hyper-parameter selection, performance was measured on the validation set. Resampling and loss-stages used their default values as shown in Table \ref{tab:stages_searchspace}. The reason for optimizing these components first were: (a) the network architecture is the central and only obligatory part of the pipeline and thus of primary importance for optimization, and (b) we hypothesized interaction effects between denoising and network architecture parametrization, since the denoising setting altered the dimensionality and nature of the input features.

\textbf{Round 2: Optimization of loss type + resampling}. For each loss variant, we investigated each possible combination of over \& undersampling according to Table \ref{tab:stages_searchspace}. Network architecture and spectrogram denoising were fixed to the optimal configuration found in round 1. We investigated these stages in combination, since we suspected interaction effects, e.g. loss variants specialized in balancing stages might prefer less oversampling. For this round we used the reduced test set for evaluation, as it more accurately mirrors the real class distribution than the validation set.

We performed all experiments separately for each class. To reduce performance variability due to random model initializations and dataset shuffling, we repeated each network training and evaluation 5 times and averaged performance results.

\section{Results}
\label{sec:results}

\subsection{Optimization of network architecture + spectrogram denoising}
\label{sec:round1_results}

\begin{figure}[t]
    \centering
    \includegraphics[width=0.4\textwidth]{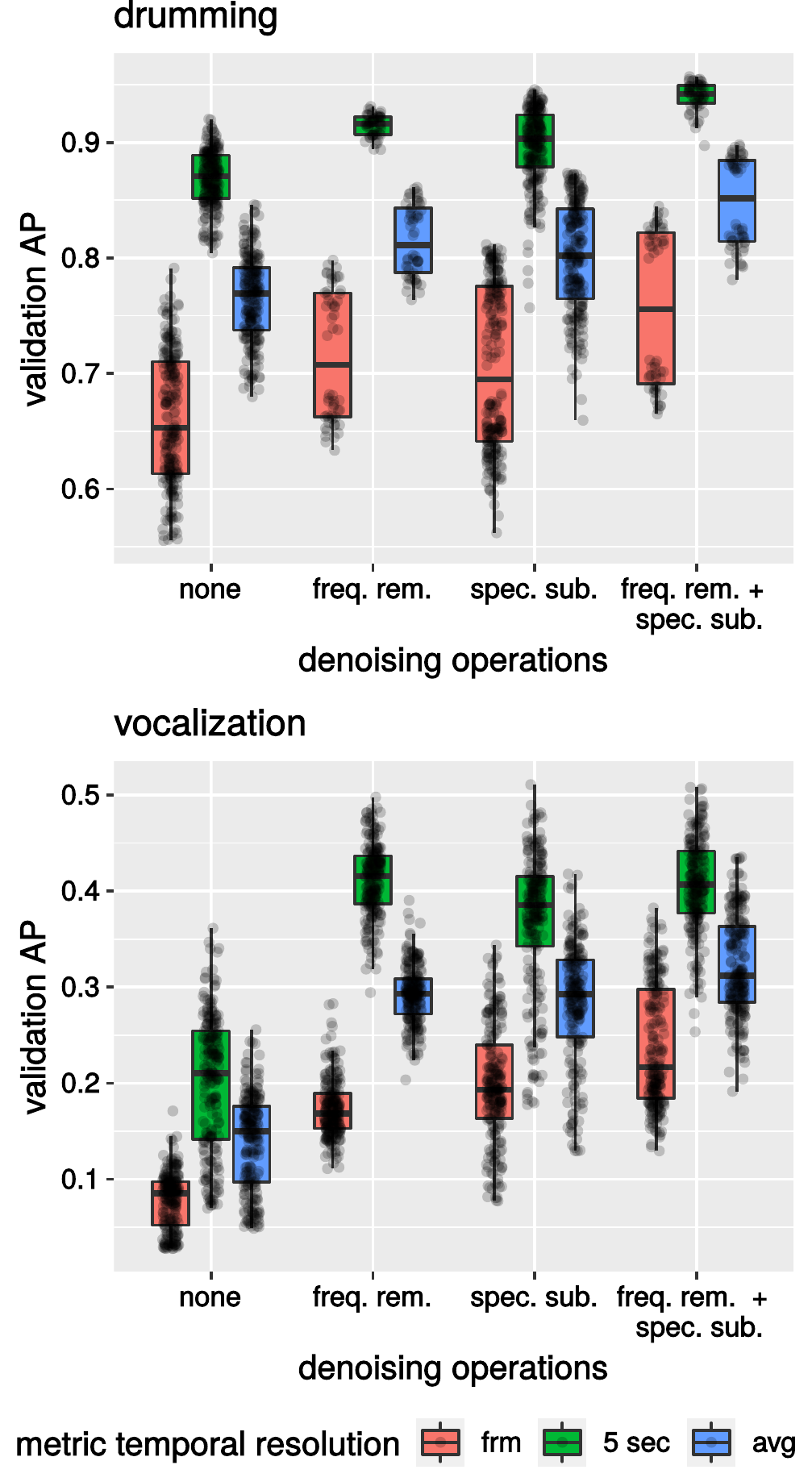}
    \caption{\textbf{Performances of network configurations, grouped by denoising operations.} The x-axis indicates spectrogram denoising operations, the y-axis indicates validation performance. Each data point presents the performance of a network configuration produced in the architecture grid search, averaged over 5 runs.}
    \label{fig:archden_valperf_overview}
\end{figure}

\begin{figure*}[t]
    \centering
    \includegraphics[width=1\textwidth]{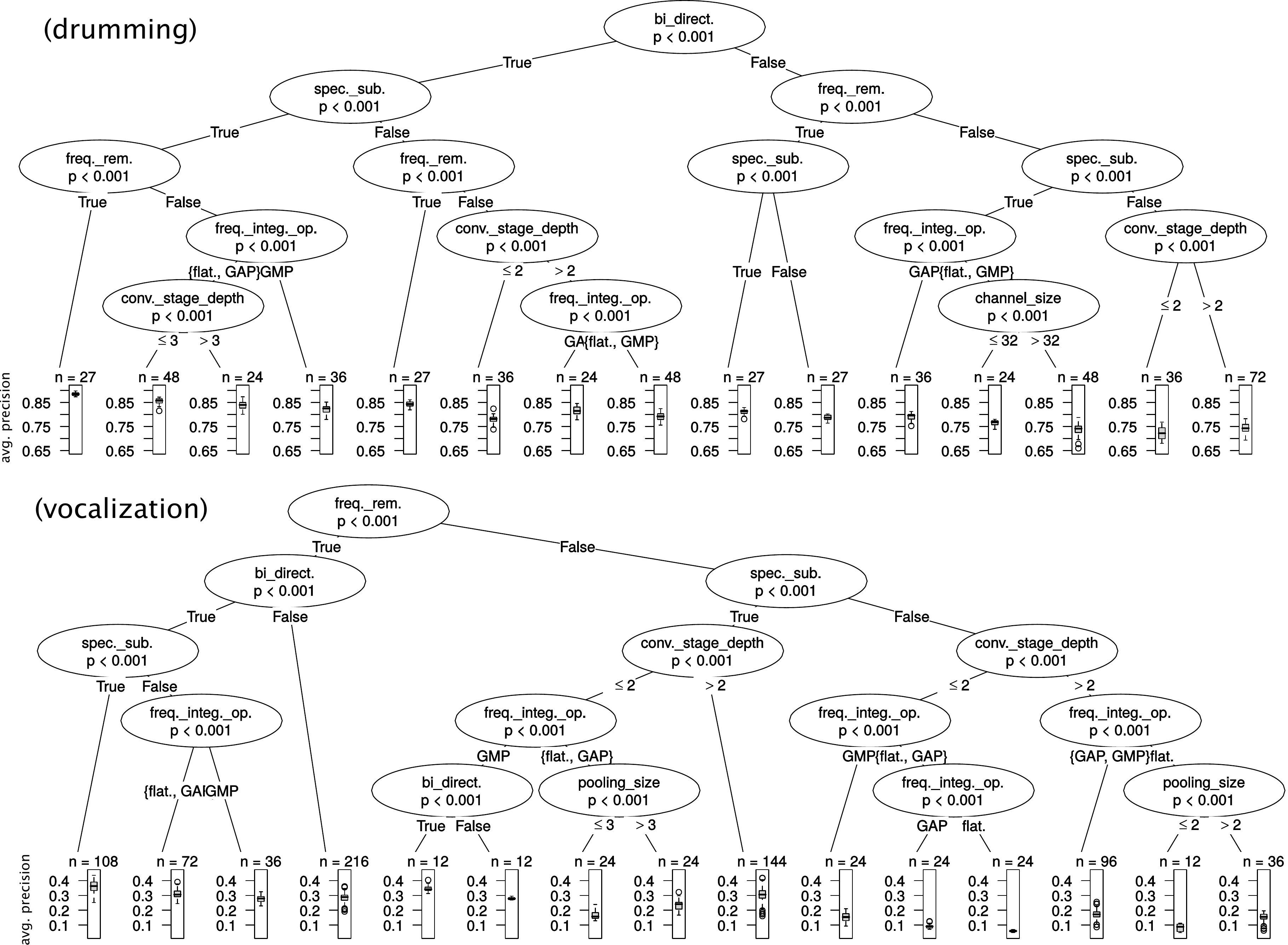}
    \caption{\textbf{Conditional inference trees for optimization of network architecture + denoising} Nodes show hyper-parameters with the p-value resulting from splitting the feature into groups indicated by the branches. Leafs show the performance distributions. }
    \label{fig:arch_ctrees_round1}
\end{figure*}

Figure \ref{fig:archden_valperf_overview} shows the validation set performances achieved by networks in the architecture grid search, grouped by denoising operations. To statistically assess the influence of the investigated parameters, we constructed conditional inference trees (c-trees), shown in Figure \ref{fig:arch_ctrees_round1}. C-trees are regression trees where the splitting criterion is the statistical significance (p-value). At each node, the tree splits instances into two subgroups based on a singular feature, choosing the split with the highest significance, until no significant split can be made. The advantage of c-trees for statistical analysis are that (1) they indicate hyper-parameter importance, as the globally most important hyper-parameters occur higher to the root, and (2) they implicitly account for interaction effects, i.e. hyper-parameter effects only occurring in subgroups. We fit regression trees on the validation performance $AP_{\text{avg}}$ with all denoising and architectural hyper-parameters as predictors. The p-value-cutoff was $0.001$ with Bonferroni-correction to limit trees to the most essential effects. C-trees were implemented with R-package \texttt{party v 1.4-5} \cite{hothorn2015ctree}.

\begin{table*}[t]
\renewcommand{\arraystretch}{1.2}
\footnotesize
\caption{\textbf{Network architectures with highest validation performance per denoising setting.} The underlined models were selected for subsequent experiments regarding loss functions and resampling. "Complete test performance" means that the complete test set was used for testing (and not the reduced test set, see Table \ref{tab:dataset}). Bold numbers highlight the highest performance reached for each class.}
\label{tab:best_models_first_opt_stage}
\centering
\begin{tabular}{@{}L{0.8cm}|L{1cm}L{1cm}|L{1cm}L{1cm}L{1cm}L{1cm}L{1cm}|C{0.7cm}C{0.7cm}C{0.8cm}|C{0.7cm}C{0.7cm}C{0.8cm}@{}}
\toprule
class                  & \multicolumn{2}{c}{denoising setting} & \multicolumn{5}{c}{architectural hyperparameters}                         & \multicolumn{3}{c}{val performance}                        & \multicolumn{3}{c}{complete test performance}                       \\
                       & freq. rem.       & spec. subtr.       & conv. stage depth & pooling size & channel size & freq. int. & bi-direct. & $AP_{\text{frm}}$ & $AP_{\text{5sec}}$ & $AP_{\text{avg}}$ & $AP_{\text{frm}}$ & $AP_{\text{5sec}}$ & $AP_{\text{avg}}$ \\ \midrule
\multirow{4}{*}{drum.} & False         & False      & 4                 & 2            & 96           & GAP         & True           & 0.791      & 0.901    & 0.846      & 0.025       & 0.059     & 0.042       \\
      & True          & False      & 2                 & 2            & 32           & GMP         & True           & 0.794      & 0.928    & 0.861      & 0.084       & 0.214     & 0.149       \\
      & False         & True       & 4                 & 2            & 96           & GAP         & True           & 0.808      & 0.939    & 0.873      & 0.044       & 0.079     & 0.061       \\
      & \underline{True}          & \underline{True}       & \underline{2}                 & \underline{2}            & \underline{96}           & \underline{GAP}         & \underline{True}           & 0.839      & 0.957    & \textbf{0.898}      & 0.162       & 0.294     & \textbf{0.228}       \\  \midrule 
\multirow{4}{*}{voc.}  & False         & False      & 3                 & 3            & 64           & GAP         & True           & 0.171      & 0.34     & 0.256      & 0.008       & 0.013     & 0.01        \\
      & True          & False      & 2                 & 3            & 96           & GAP         & True           & 0.283      & 0.498    & 0.39       & 0.015       & 0.02      & \textbf{0.018}       \\
      & False         & True       & 3                 & 3            & 32           & GMP         & True           & 0.325      & 0.511    & 0.418      & 0.011       & 0.014     & 0.013       \\
      & \underline{True}          & \underline{True}       & \underline{2}                 & \underline{4}            & \underline{96}           & \underline{GAP}         & \underline{True}           & 0.382      & 0.488    & \textbf{0.435}      & 0.016       & 0.02      & \textbf{0.018}      \\
 \bottomrule      
\end{tabular}
\end{table*}

We highlight the following observations based on these visualizations:

(1) Detection performance was drastically higher for drumming than for vocalization, regardless of network architecture and denoising setting (drumming $AP_{\text{avg}}$ range: .55 - .95, vocalization: .03 - .50). Drumming performances reached up to almost perfect scores, while the worst drumming performance was still higher than the best vocalization performance.

(2) $AP_{5}$ values were generally higher than $AP_{\text{frm}}$ values for both target classes. This means that networks were better at detecting rough localization of target events than frame-precise localization (see Fig. \ref{fig:archden_valperf_overview}). 

(3) Both spectrogram denoising operations increased performance $AP_{\text{avg}}$ on average (i.e. over all network configurations constructed in the grid search). Using no denoising operation yielded significantly ($p<0.0001$) lower performance than using either or both operations in combination. Using both operations simultaneously yielded a significantly higher ($p<0.0001$) performance than using either or none. The performance difference between using either operation individually was not significant ($p>0.01$).

(4) Both spectrogram denoising operations increased performance regarding the maximum $AP_{\text{avg}}$ reached by a model constructed in the architecture grid search. The order was no preprocessing $\mapsto$ freq. rem. $\mapsto$ spec. sub $\mapsto$ freq. rem. + spec. sub for both classes.

(5) The usage of a bi-directional recurrent layer was the most important network architecture property, i.e. it increased performance for both classes with most denoising operations. For vocalization, they only increased $AP_{\text{frm}}$ and not $AP_{\text{5}}$, i.e. they only increased the capability for precise localization of target events while rough allocation remained the same. The two clusters in Fig. \ref{fig:archden_valperf_overview} for drumming with spec. sub. + freq. rem. are explained through bi-directional layers.

(6) The influence of network architecture choices decreased when applying integration operations. This is evidenced by the fact that both c-trees (Fig. \ref{fig:arch_ctrees_round1}) have the largest depths in the paths without denoising operations. The network features with the most influence were the depth and the choice of frequency integration operation. However, which depth and integration operation optimized performance was dependent upon the denoising setting. On average, performance increased with depth and global average pooling for frequency integration.

Table \ref{tab:best_models_first_opt_stage} shows architectures with the highest validation performance $AP_{\text{avg}}$ per denoising setting and their test set performances. We highlight the following observations:

(1) Test performances dropped drastically compared to the validation set. This is largely due to the test set containing far more negative examples than the validation set (see Table \ref{tab:dataset}).

(2) While frequency removal and spectral subtraction performed similarly on the validation set, frequency removal outperformed spectral subtraction on the test set. Combining both operations yielded the highest test performance for drumming, and performed on par with using frequency removal alone for vocalization.

(3) For drumming, the importance of denoising functions increased on the test set. The performance difference between using no and both operations was .05 on the validation set, but .18 on the test set. For vocalization, denoising importance decreased on the test set. The performance difference between using none and both was .28 on the validation set, but only .08 on the test set.

For the experiments on loss function + resampling in section \ref{sec:round2_results} we used the setting with the highest validation performance $AP_{\text{avg}}$ for each class, i.e. the underlined models in Table \ref{tab:best_models_first_opt_stage}.

\subsection{Optimization of Loss Variant + Resampling}
\label{sec:round2_results}

\begin{figure*}[t]
    \centering
    \includegraphics[width=0.8\textwidth]{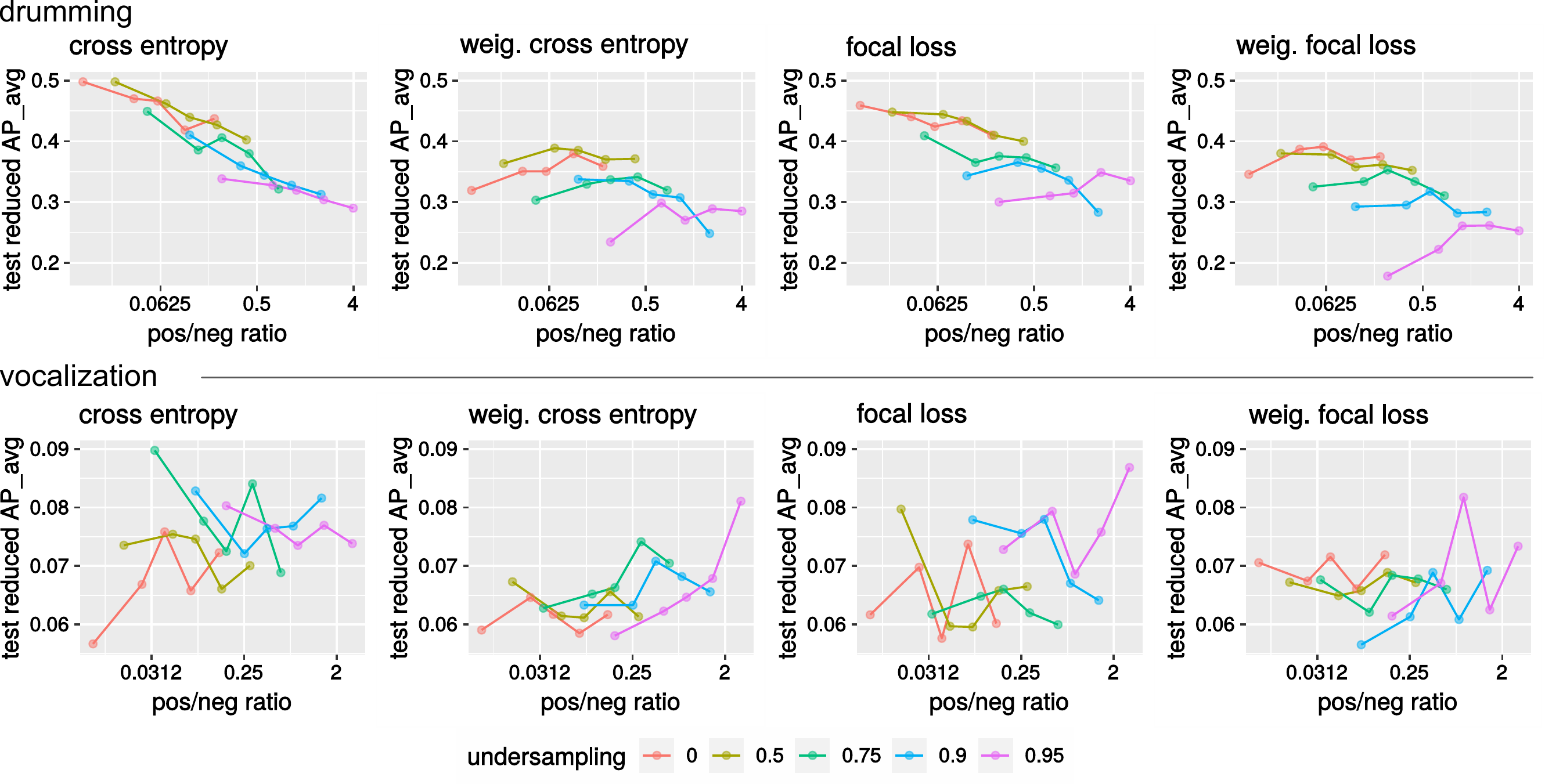}
    \caption{\textbf{Influence of resampling grouped by loss functions} Upper row: drumming, lower row: vocalization. The x-axes indicate the the ratio of positive to negative segments. Oversampling is shown implicitly, where 5 data points per undersampling setting correspond to oversampling duplication amounts $\{ 0,2,4,8,16 \}$.}
    \label{fig:loss_vs_resampling}
\end{figure*}

\begin{figure}[t]
    \centering
    \includegraphics[width=0.35\textwidth]{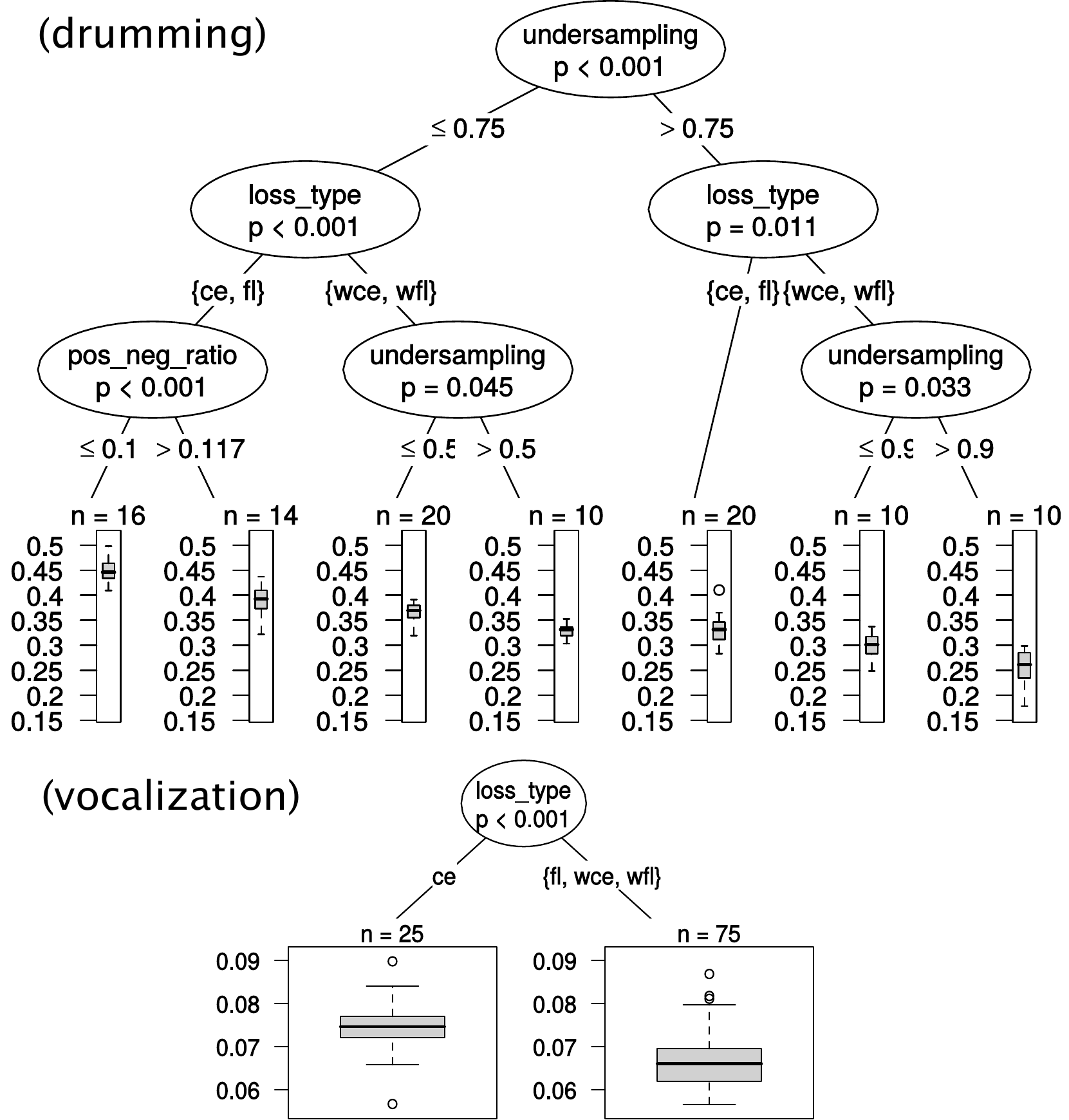}
    \caption{\textbf{Conditional inference trees for loss type and resamling}}
    \label{fig:ctrees_losstype_resampling}
\end{figure}

Figure \ref{fig:loss_vs_resampling} shows the results of the experiments on the loss variant and resampling, measured on the reduced test set as $AP_{\text{avg}}$. Additionally, we calculated the total ratio of positive to negative segments resulting from the over / undersampling settings (positive segment = segment with at least one positive frame, see sec. \ref{sec:resampling}). Figure \ref{fig:ctrees_losstype_resampling} shows the corresponding c-tree analysis analogous to the one of section \ref{sec:round1_results} (target: reduced test set $AP_{\text{avg}}$, predictors: loss variant and resampling parameters + pos/neg ratio).

Loss and resampling had greater influence on drumming than vocalization, i.e. performance range was .2 - .5 for drumming and .05 - .09 $AP_{\text{avg}}$ for vocalization. For drumming: 

(1) Both unweighted loss functions reached higher performances than weighted functions. The global effect of weighted vs unweighted functions was significant ($p < 0.001$). The global difference between unweighted cross entropy and focal loss was not significant. However, standard cross entropy had significantly higher performance than focal loss ($p < 0.01$) when undersampling $\leq 0.5$. 

(2) Performance decreased globally with increased undersampling.

(3) For unweighted loss functions there was a significant global association for increased performance with a decreased ratio of positive/negative segments ($p < 0.001$). This also means that performance decreased with increased oversampling. 

(4) The highest performance $AP_{\text{avg}} = .49$ was reached with binary cross entropy using the "raw" training database without any resampling.

For vocalization, the only significant influence was standard binary cross entropy performing better than the other loss functions, although the influence was still low in absolute terms. Resampling had no systematic influence. The highest value $AP_{\text{avg}} = .09$ was reached with no oversampling and undersampling of $0.75$.

Table \ref{tab:final_eval} shows the final performance evaluation on the complete test set for the best hyper-parameter settings found in this study. The denoising settings and network architecture setting correspond to the settings underlined in Table \ref{tab:best_models_first_opt_stage}. The loss type and resampling correspond to the settings with the highest performance found on the reduced test set (see figure \ref{fig:loss_vs_resampling} ). $AP_{\text{avg}}$ values dropped by approximately \SI{60}{\percent} from the reduced to the complete test set. This is due to the sevenfold increase of negative examples compared to the complete test set, increasing the amount of possible false positive predictions.

The baseline performance corresponds to the performance achieved by Heinicke et al. \cite{heinicke2015assessing}. Their $F1$ values were computed on event-based metrics with varying event lengths based on the segmentation algorithm in their study, i.e. they are not directly comparable to our segment-based metrics. When comparing the baseline values to $F1_{\text{avg}}$-values, our performances were an improvement, increasing baseline performance. For drumming the increase was $\SI{30}{\percent} F1$, which is a 7-fold increase. For vocalization, the increase was $\SI{5}{\percent} F1$, which is a 25-fold increase.

\begin{table*}[t]
\renewcommand{\arraystretch}{1.2}
\footnotesize
\caption{\textbf{Final performane evaluation}}
\label{tab:final_eval}
\centering
\begin{tabular}{@{}L{1cm}| L{1cm}|L{1cm}L{1cm}L{1cm}L{1cm}|L{1cm}L{1cm}L{1cm} | L{1cm}L{1cm}L{1cm}| L{1cm}@{}}
\toprule
class & loss  & \multicolumn{2}{c}{resampling} & \multicolumn{2}{c}{pos/neg ratio train set} & \multicolumn{7}{c}{complete test set performance}                                                                 \\ 
                                                               &                        & undersamp.  & oversamp.  & frm                  & seg                 & $AP_{\text{frm}}$ & $AP_{\text{5}}$ & $AP_{\text{avg}}$ & 
                                                               $F1_{\text{frm}}$ & $F1_{\text{5}}$ & $F1_{\text{avg}}$ & baseline $F1$ \cite{heinicke2015assessing} \\ \midrule
drum.                                                                                          & cross entr.                                  & 0              & 0             & 0.002                & 0.012               & 0.308       & 0.385     & 0.347       & \SI{34}{\percent}       & \SI{32.6}{\percent}     & \SI{33.3}{\percent}       & \SI{4.6}{\percent}       \\
voc.                                                                                                    & cross entr.                                  & 0.75           & 0             & 0.011                & 0.034               & 0.018       & 0.025     & 0.021       & \SI{5}{\percent}        &\SI{5.3}{\percent}     &\SI{5.1}{\percent}   & \SI{0.2}{\percent}       \\ \bottomrule
\end{tabular}
\end{table*}

\section{Discussion}
\label{sec:discussion}

The F1 scores of $\SI{33}{\percent}$ for drumming and $\SI{5}{\percent}$ for vocalization might seem rather low in absolute terms. However, one has to take into account that even for humans the problem is exceptionally difficult. In addition to the rarity of the target calls, they are also very faint and subtle. This issue requires extensive training by human listeners to label calls reliably. Hence, although our results leave room for further improvement, they represent an improvement compared to previous methods.

The detection performance for chimpanzee drumming was drastically higher than for vocalization, with and without denoising/resampling. The same effect occurred in Heinicke's \cite{heinicke2015assessing} method, although absolute values were lower. We hypothesize that the following factors contribute to the difficulty of detecting chimpanzee vocalizations: (1) Vocalizations are more complex with greater intra-class variability than drumming, as they encompass multiple call types with respect to pant hoots and screams. In comparison, drumming has a more "fixed" and stereotypical pattern. (2) The frequency bands for chimpanzee vocalization are also occupied by calls of other primate species which are acoustically similar. However, drumming is the only animal call occupying such low frequency bands in our data set. In our experience, human listeners also have greater difficulty in identifying chimpanzee vocalizations, particularly because they confuse them for other animal calls. Thus we reason that the vocalization class may have needed more training examples to be learned effectively by the network, given the greater difficulty of the task.

Both denoising operations, spectral subtraction and frequency removal, increased performance significantly, particularly for drumming's test set performance. This finding is in accordance with other studies on animal call detection which applied similar operations with success \cite{xie2020bioacoustic, mac2018bat, himawan2018deep}. We were particularly surprised by the magnitude of increase in performance by frequency removal. Theoretically, networks should learn to ignore irrelevant frequency bands by themselves. We give two possible explanations for this: (1) Positive class examples were only present for few recordings. Possibly, this led the network to infer a false association between noise conditions and target call occurence. Removing uncorrelated frequencies reduced the features which could be used for such false associations. (2) Possibly, the test set contained background noise conditions which were drastically different from the ones in the training set so that they occupy regions far away from the learned manifold and cause faulty forward-passes in the network, similar to adversarial examples \cite{smith2018understanding}. We highlight that frequency removal carries the additional advantage of decreasing computation time.

We found that taking relative class balance did not increase performance. Drumming reached the highest performances using standard binary cross entropy loss without any data set resampling, i.e. using the raw, heavily imbalanced training data. Vocalization also performed best with vanilla cross entropy, but was insensitive to resampling. We draw the conclusion that performance-wise, combating relative class imbalance is unnecessary or even harmful. For drumming, undersampling decreased the performance regardless of the loss function, i.e. displaying diversity of the background class is important even if examples might seem redundant for humans. Still, undersampling can reduce training time with minimal loss in performance if used only slightly. Drumming reached essentially the same performance using \SI{0}{\percent} and \SI{50}{\percent} of the training data and began dropping when only using \SI{25}{\percent}. This finding is in contrast to other studies \cite{ lin2017focal, weiss2004mining, haixiang2017learning, johnson2019survey}, which usually report positive effects for balancing methods. We give the following possible explanations for this discrepancy: (1) Studies reporting positive effects of resampling commonly worked with imbalanced training sets, but balanced test sets \cite{buda2018systematic, hensman2015impact}. Consequently, the positive effect of resampling could be attributed to approximating class distributions between training and test set, and not to compensating the imbalance within the training set. (2) Studies reporting positive effects of class weights in classification settings usually performed multi-class single-label classification with softmax activation as in multinominal logistic regression \cite{king2001logistic, wang2018predicting, anand1993improved}. However, we used binary cross entropy for single-class prediction, which might be inherently more robust to class imbalances. (3) When Lin et al. reported focal loss to outperform binary cross entropy, they performed multi-label detection \cite{lin2017focal} for 91 classes with one network for the COCO dataset. As the influence of background class multiplicates across all positive classes in multi-label detection, loss balancing might become beneficial in such multi-label settings.

In summary, our results show that supporting the network to learn decoupling target class characteristics from background class characteristics is of primary importance for increasing performance. Spectrogram denoising explicitly supports this decoupling by discarding information from signals which are assumed to only be associated with background noise based on prior knowledge. Including more examples from the background class (no undersampling) implicitly supports this decoupling by displaying a greater amount of background noise variability to the network.

\section{Conclusion and Future Work}
\label{sec:conclusion}

In this paper we investigated the automatic detection of chimpanzee drumming and vocalizations in long-term forest recordings of PAM. The detection approach was based on convolutional recurrent neural networks with spectrogram inputs. The particular challenge of this task was the severe class imbalance and rarity of target calls. We applied various extensions to the pipeline for compensating this imbalance: We evaluated two spectrogram denoising operations frequency removal (i.e. removing frequency bands outside the target call's range) and spectral subtraction. Both operations significantly improved the performance. For mitigating relative class imbalance, we evaluated various loss functions (weighted / unweighted cross entropy / focal loss) as well as random over/undersampling of segments. The best performing loss was unweighted binary cross entropy. For drumming, any resampling decreased performance, i.e. training on heavily imbalanced training data reached the highest performance. For vocalization, resampling had no significant effect. From this we conclude that a primary factor for increasing performance in animal call detection in PAM settings is aiding the network to learn decoupling background noise conditions from target call characteristics. Final performance results were an improvement on the previous baseline performance.

With an algorithm performance of about \SI{30}{\percent} for the detection of chimpanzee drumming, this approach may become suitable for continuous field monitoring. As previously demonstrated, species occurrence \cite{kalan2015towards} and movement patterns \cite{kalan2016passive} can be modeled using PAM data. Such applications may be enhanced using the detection process proposed in this study.

We see the greatest chances for further improvement by investigating the following areas: 

(1) Applying methods for increasing diversity in the existing positive examples. The most straight-forward method is data augmentation, i.e. applying perturbations to the positive calls to diversify their patterns. Particularly mix-up could prove successful to further teach decoupling of noise and positive examples \cite{zhang2017mixup}, i.e. randomly overlapping positive segments with noise samples. 

(2) Increasing the amount of positives examples. This could be done by applying the algorithm to the yet unlabeled data and collecting additional true positives. Alternatively, one could generate synthetic examples through generative adversarial networks (GANs).

(3) Methods for informed undersampling such as SMOTE. Multiple studies show that informed undersampling can outperform random undersampling  \cite{weiss2004mining, haixiang2017learning, johnson2019survey}. Even if performance could not be further improved through undersampling, it may be possible to reduce training time by carefully selecting a representative subset of negative examples.

\section*{Acknowledgment}

This work was funded by the research grant for doctoral researchers at Leipzig University of applied sciences, grant number 3100451136, as well as the European Union as part of the ESF-Program, grant number K-7531.20/434-11; SAB-Nr. 100316843. We thank the Minist\`{e}re de la Recherche Scientifique, the Minist\`{e}re de l'Environnement et des Eaux et For\^{e}ts and the Office Ivorien des Parcs et Reserves of C\^{o}te d'Ivoire for permissions to work in the country and Ta\"{i} National Park.



\end{document}